\documentclass{ws-procs9x6}

\newcommand{\be}{\begin{equation}}
\newcommand{\ee}{\end{equation}}
\newcommand{\te}{t_{\rm E}}
\newcommand{\re}{\theta_{\rm E}}
\newcommand{\vt}{\mu_{\rm rel}}
\newcommand{\sm}{\mbox{M}_{\odot}}
\newcommand{\tfw}{t_{1/2}}

\newcommand{\ga}{\;\raisebox{-.8ex}{$\buildrel{\textstyle>}\over\sim$}\;}
\newcommand{\ts}{\textstyle}

\begin{document}

\title{Latest results from the POINT-AGAPE pixel-lensing survey of the
Andromeda Galaxy}

\author{Eamonn Kerins}

\address{Astrophysics Research Institute,\\
	Liverpool John Moores University,\\
	12 Quays House, Egerton Wharf,\\
	Birkenhead CH41 1LD, UK.\\
E-mail: ejk@astro.livjm.ac.uk}


\maketitle

\abstracts{I report on recent results from the POINT-AGAPE
pixel-lensing experiment, which is engaged in a search towards the
Andromeda galaxy (M31) for gravitational microlensing signatures from
massive compact halo objects (Machos). An analysis of two years of
data reveals over 360 light-curves compatible with microlensing. The
third year of data, currently being analysed, will be crucial in
determining how many of these candidates are long-period variables
rather than microlensing. Within the dataset we have isolated a subset of
four high signal-to-noise ratio, short duration events which are
compelling microlensing candidates. The properties and possible
origins of these events are discussed.
}

\section{Microlensing and pixel lensing}

For about a decade a number of microlensing surveys have focused on
the Magellanic Clouds to look for evidence of dark matter in the form
of massive compact halo objects (Machos) in our own
Galaxy\cite{mic}. They have monitored millions of resolved stars on a
nightly basis in order to detect rare transient magnifications of
starlight due to the passage of intervening objects. The two key
microlensing observables are the Einstein radius crossing time, $\te
\propto m^{1/2}$, where $m$ is the Macho mass, and the microlensing
rate, $\Gamma \propto f m^{-1/2}$, where $f$ is their fractional
contribution to the dark matter density. Galactic microlensing surveys
determine $\te$ directly from the light-curve of the event:
   \be
	F(t) = A(t) F_* = \frac{u^2 + 2 }{u\sqrt{u^2 +4}} F_*; 
	\quad
	u^2 = u_0^2 + \left( \frac{t - t_0}{\te} \right)^2,
	\label{nondeg}
   \ee
where $F$ is the flux observed at epoch $t$, $A$ is the microlensing
magnification, $F_*$ is the baseline flux of the source star, and $u_0
\equiv u(t_0)$ is the impact parameter in units of the Einstein
radius.

More recently a new technique, pixel lensing, has allowed microlensing
surveys to target more distant galaxies in which most of the sources
are unresolved. For such galaxies microlensing is detected as a small
enhancement in pixel flux within a seeing disk. A detection
typically requires high magnification $(A \ga 10)$, when
the excess flux due to lensing is $(A-1)F_* \simeq F_*/u$.
Pixel-lensing light-curves therefore take on a degenerate form\cite{gou96}:
   \be
	F(t) = F_{\rm bg} + (A-1)F_* \simeq F_{\rm bg} +
	\frac{\Delta F_0}{\sqrt{1+\left( \frac{\ts t-t_0}{\ts u_0
	\te} \right)^2}} \label{deg},
   \ee
where $F_{\rm bg}$ is the contribution to the seeing disk flux from
the local galactic surface brightness (unresolved stars) and sky
background, and $\Delta F_0 = F_*/u_0$ is the excess flux from the
lensed source at maximum magnification. Hence, instead of $\te$, the
observed timescale is the full-width at half-maximum duration, $\tfw
\simeq 2\sqrt{3} u_0 \te$. Therefore, with the exception of very high
signal-to-noise ratio (S/N) events, one cannot disentangle $F_*$,
$u_0$ and $\te$ in pixel lensing.

An additional complication is that seeing and sky background
variations must be minimised to reveal real flux variations. Over the
last six years methods have been developed\cite{dimage} to
accomplish this and real variations at the level of $\sim 1\%$ can now
be detected reliably. The power of pixel lensing rests in the fact
that large external galaxies can be probed, potentially providing
large numbers of Macho events and allowing their distribution to be
mapped across the target galaxy. The information this provides makes
up for the the loss of timescale information.

\section{The POINT-AGAPE survey}

\begin{figure}[ht]
{\epsfysize=2.25in\epsfbox{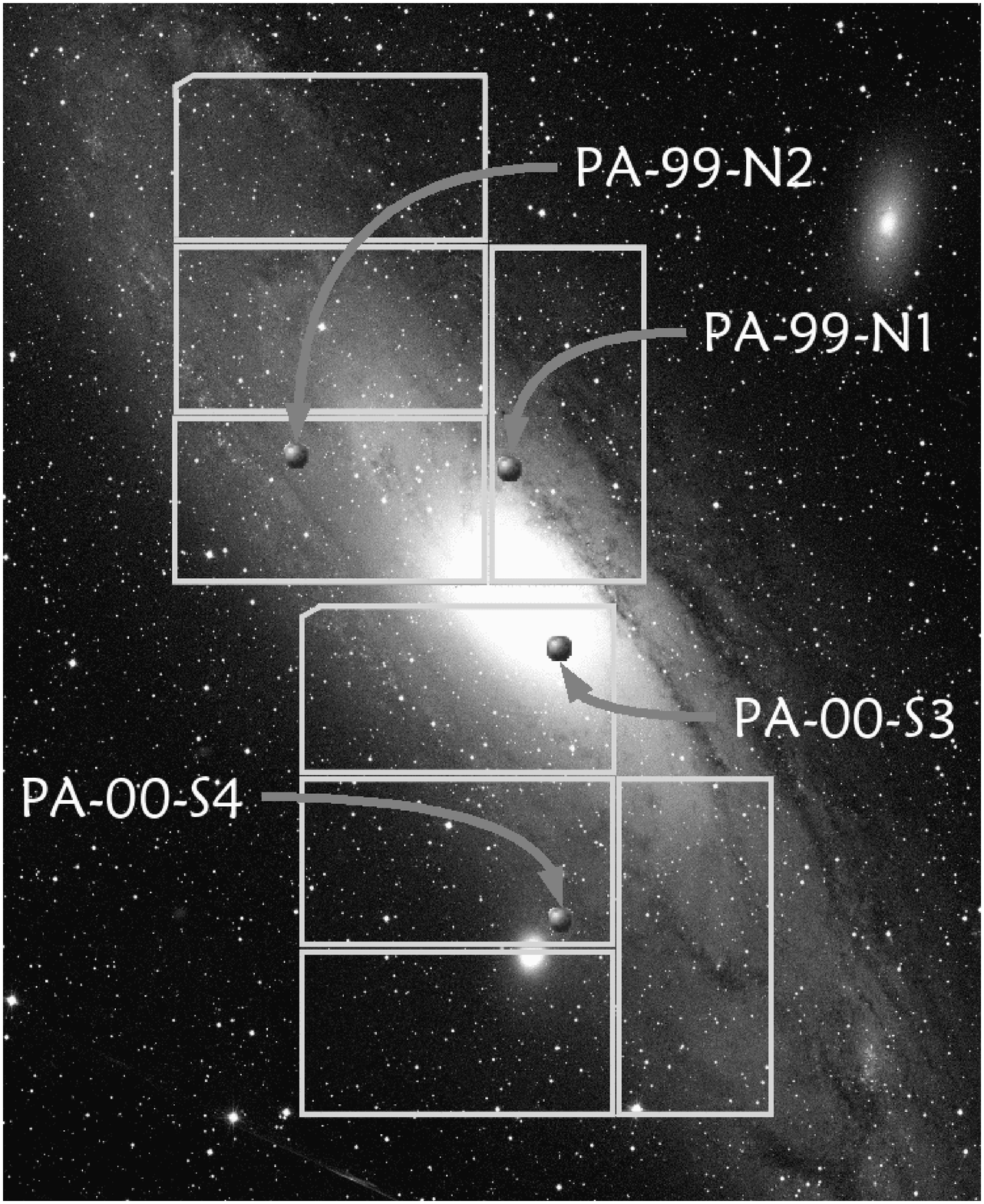}\hspace{0.6cm}\epsfysize=2.in\epsfbox{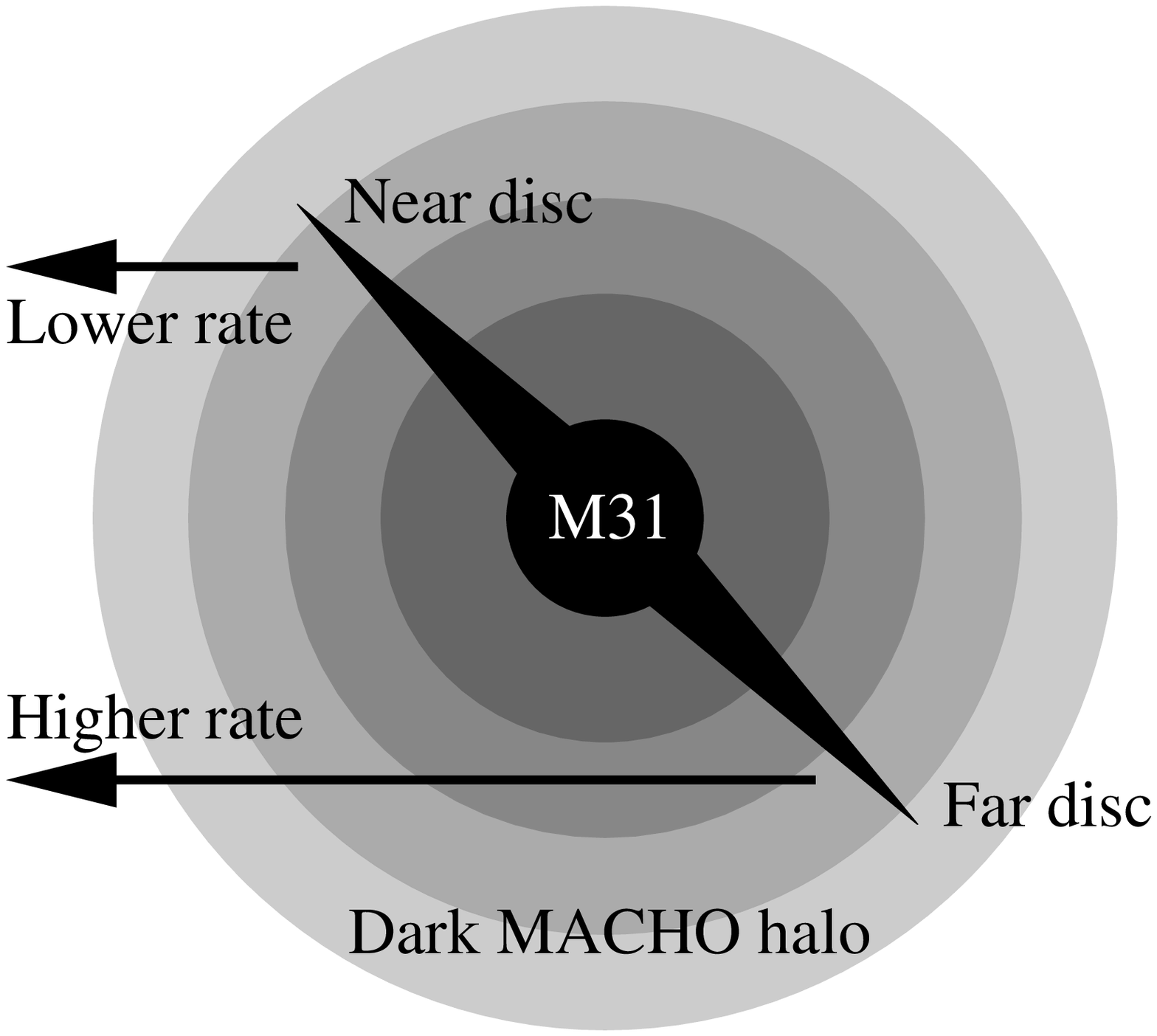} }
\caption{(Left) The two POINT-AGAPE fields showing the positions of the
four high S/N, short-duration candidate events. (Right) A schematic
representation of M31 showing how the inclination of the M31 disk
produces a gradient in the Macho microlensing rate from the near to
far disk.  (M31 image: Bill Schoening, Vanessa Harvey/REU
programme/AURA/NOAO/NSF.)
\label{location}}
\end{figure}

The POINT-AGAPE survey is an Anglo-French experiment which is
targeting M31 using the wide-field camera (WFC) on the 2.5~m Isaac
Newton Telescope (INT). POINT-AGAPE is the successor to the pilot
AGAPE (Andromeda Galaxy Amplified Pixels Experiment) survey, where
POINT is an acronym for ``Pixel-lensing Observations with INT''. Since
1999 we have been observing a large region of the M31 disk (left-hand
panel of Fig.~\ref{location}) in the Sloan-like $r'$ band, augmented
by observations in either Sloan-like $i'$ or $g'$. Our exposures are
typically 5-10 mins per night and we have been able to collect data on
most nights, weather permitting, when the WFC is mounted during the
M31 observing season (August to January). To date we have around 180
epochs of data spanning three seasons.

The principal goal of the survey is to measure or constrain the
abundance of Machos from the Jupiter mass scale up to several Solar
masses. M31 provides a powerful signature for Macho detection
(right-hand panel of Fig.~\ref{location}). Its high disk inclination
$(i = 77^{\circ})$ should produce an asymmetry in the M31 Macho
spatial distribution if they occupy a spheroidal dark
halo\cite{cro92}. This permits M31 Machos to be discriminated
statistically from Milky Way Machos, and from stellar microlenses or
variable stars in the M31 disk. Combining this spatial information
with the number of events and their $\tfw$ duration allows the M31
Macho mass and density contribution to be measured\cite{ker01}, though
we expect that at least three seasons of data are required. Milky Way
Machos can also be detected, though the lack of an obvious spatial
signature makes their contribution more difficult to measure.

\section{Current status}

We have analysed the first two seasons of data, and are currently
analysing our third season. We have defined a set of event selection
criteria based upon the goodness of fit to a theoretical microlensing
curve, and upon our sensitivity, sampling and observation
baseline. These criteria have produced 362 candidate
events\cite{pau02b}. With only two seasons of data we are not yet in a
position to eliminate some classes of long-period variables, such as
Miras, and we suspect many of our candidates are not microlensing
events. A more accurate calibration of the number of microlenses must
therefore await completion of the third-year analysis.

\begin{figure}[ht]
\centerline{\epsfxsize=4.5in\epsfbox{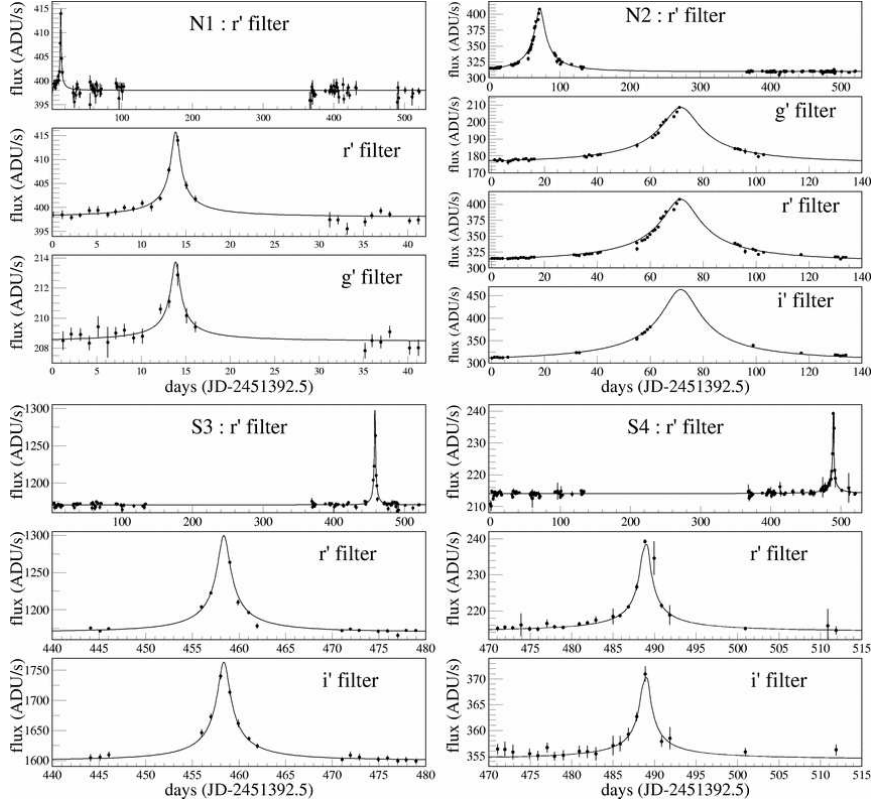}}   
\caption{Light-curves in $g'$, $r'$ and $i'$ bands of the four high
S/N, short-duration candidate events. Clockwise from top-left the
durations $(\tfw,\te)$ of the events in days are: $(1.9,9.7\pm0.7)$,
$(25,92\pm4)$, $(2.1,129^{+143}_{-72})$, and $(2.3,13^{+5}_{-3})$.
\label{candidates}}
\end{figure}
In the meantime we have isolated a small subset of four high S/N
events, which are compelling microlensing candidates\cite{pau02b}. 
All have $\tfw < 25$~days and $\Delta F_0$ corresponding to a $R
\leq 21$~mag star. The high S/N and short $\tfw$ together provide a
very high level of discrimination against bright, long-period variable
stars. The positions of the four events are indicated in the left
panel of Fig.~\ref{location}, and the multi-colour light-curves are shown in
Fig.~\ref{candidates}. Each candidate provides a very good fit to
microlensing in all bands.
\begin{figure}[ht]
\centerline{\epsfxsize=3.5in\epsfbox{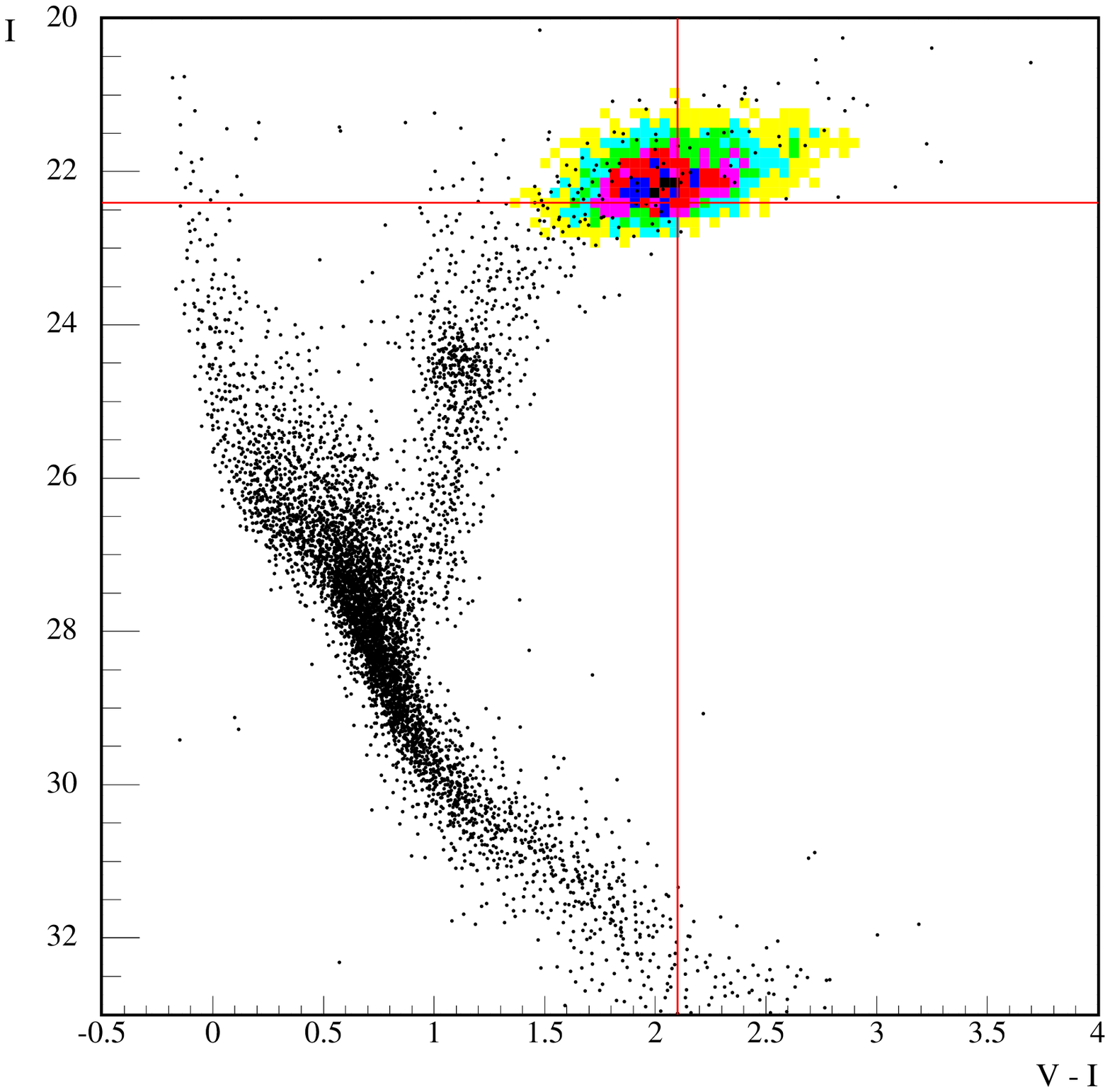}}   
\caption{Superposition of the Hipparcos colour magnitude diagram
(black points) and M31 stars resolved by HST (pixelated area). The
Hipparcos stars have been moved to the distance modulus of M31. The
colour and location of the source star for PA-99-N1 has been
determined directly from our INT data. An object consistent with these
measurements was found on HST archive data. The horizontal and
vertical lines show the colour and magnitude of the HST
object.\label{hst}}
\end{figure}

\section{Interpretation of the candidate events}

Remarkably, for three of the events we have been able to break the
light-curve degeneracy of Eqn.~(\ref{deg}) and measure, or strongly
constrain, $\te$. In the case of PA-99-N1 this has been achieved
because the source star has almost certainly been identified on
archive Hubble Space Telescope (HST) images\cite{aur01}. The position
and colour of the event as measured from INT data provide a good match
to a resolved giant star on the HST frames (see Fig.~\ref{hst}). Since
$\te = \re/\vt$, where $\re \propto m^{1/2}$ is the angular size of
the Einstein radius and $\vt$ is the relative proper motion of the
lens across the line of sight, a measurement of $\te$ leads to a
constraint on the lens mass. The small value of $\te = 9.7$~days for
PA-99-N1 indicates that either $m$ is small or else $\vt$ is
large. The excellent fit of this event to a point-source light-curve
indicates that $\re$ must be much larger than the angular size of the
source star, and requires that $\vt > 0.3\,\mu$as/day. If PA-99-N1 is
a Macho then it is likely to be a brown dwarf, but if it is a star in
M31 the strong proper motion constraint demands a lens mass in the
restricted range $m = 0.27^{+0.21}_{-0.12}\,\sm$, making it a typical
M dwarf star.

For two other events, PA-99-N2 and PA-00-S3, their high S/N is
sufficient to provide a measure of $\te$. Microlensing events are
expected to have a broad timescale distribution centred on $\te \sim
(50,60,30,20) \times (m/\sm)^{1/2}$ days for (Macho, disk--disk,
bulge--bulge, bulge--disk) lensing, respectively. PA-00-S3 is almost
certainly due to a bulge star, however the other events may be due to
either Machos or stellar lenses. Intriguingly, the proximity of
PA-00-S4 to the satellite galaxy M32, together with the colour of the
event, suggest that this system involves a stellar lens in M32 and an
M31 source star\cite{pau02a}. If it is not a Macho then PA-00-S4 is
highly likely to be the first detected intergalactic microlensing
event.

These initial results mark an encouraging first step towards the goal
of detecting or constraining the Macho population. High S/N events in
particular constitute an important sub-sample because often their
$\te$ can be measured or strongly constrained, providing tight limits
on the masses of individual lenses. More generally, the entire
pixel-lensing dataset will allow us to probe statistically the
contribution and mass scale of Machos. Pixel-lensing experiments are
also demonstrating how galactic microlensing surveys need not be
confined to our own Galaxy, and that microlensing is now becoming a
powerful and unique tool for studying the global structure and stellar
inventory of Local Group galaxies.

\end{document}